\documentclass[conference,a4paper]{IEEEtran}%
\usepackage[utf8]{inputenc}
\usepackage{graphicx}
\usepackage{tabularx}
\usepackage{amsmath}
\usepackage{amssymb}
\usepackage{stackengine}
\usepackage{makecell}
\usepackage{multirow}
\usepackage{cite}
\usepackage{array}
\usepackage{bbding}
\usepackage{pifont}
\usepackage{wasysym}
\usepackage{comment}
\usepackage{caption}
\usepackage{subcaption}
\usepackage[noend]{algpseudocode}
\usepackage{color}
\usepackage{booktabs}
\usepackage{mathtools}
\usepackage{dirtytalk}
\usepackage{etoolbox}
\usepackage{tikz}
\usepackage{algorithm,algpseudocode}
\usepackage{url}
\usepackage{float}
\usepackage{threeparttable}
\usepackage{stfloats}
\usepackage{titlesec}
\usepackage[caption=false]{subfig}
\titlespacing*{\section}{0pt}{0.5ex}{0.5ex}

\usepackage{subcaption}

\title{Emission-Aware Optimization of Container Handling Vehicles: A HaminaKotka Port Case Study}

\makeatletter
\newcommand{\linebreakand}{%
  \end{@IEEEauthorhalign}
  \hfill\mbox{}\par
  \mbox{}\hfill\begin{@IEEEauthorhalign} \vspace{-10cm}
}
\makeatother
\author{
  \IEEEauthorblockN{Hafiz Majid Hussain}
  \IEEEauthorblockN{ \textit{School of Energy Systems   }}
  \IEEEauthorblockA{
       \textit{ Lappeenranta-Lahti } \\ \textit {University of Technology}\\
    Lappeenranta, Finland \\
    Majid.hussain@lut.fi
   }
  \and
  \IEEEauthorblockN{Juha Haakana}
\IEEEauthorblockN{ \textit{School of Energy Systems   }}
  \IEEEauthorblockA{
      \textit{ Lappeenranta-Lahti } \\ \textit {University of Technology}\\
    Lappeenranta, Finland \\
    Juha.Haakana@lut.fi  }

 \and
 \IEEEauthorblockN{Mehar Ullah}
  \IEEEauthorblockN{ \textit{School of Energy Systems   }}
  \IEEEauthorblockA{
      \textit{ Lappeenranta-Lahti } \\ \textit {University of Technology}\\
    Lappeenranta, Finland \\
    Mehar.Ullah@lut.fi
   }
} 

\makeatletter

\begin{document}  
\maketitle
\begin{abstract}
Smart ports increasingly depend on IoT-enabled digitalization to turn heterogeneous operational processes into measurable, optimizable workflows. Accordingly, this paper first provides a concise review of IoT-enabler technologies for port digital transformation, emphasizing how sensing, connectivity, and data integration capabilities support real-time decision-making in port environments. Motivated by the growing emphasis on operational decarbonization reflected in the European Green Deal and the Sustainable Development Goals (SDGs),  we  then present a data-driven fleet scheduling framework for landside port operations at the Port of HaminaKotka (Mussalo terminal), Finland. The scheduling framework reallocates vehicle operating hours to minimize total CO$_2$eq emissions while preserving the same aggregate workload. The study uses a two-year dataset from 77 port-operating vehicles, analyzed at monthly and annual aggregation levels, to derive vehicle-specific intensity metrics (CO$_2$eq per operating hour) and identify reallocation opportunities. We formulate the scheduling task as a linear programming workload allocation problem that minimizes the weighted sum of assigned hours while keeping total required hours fixed, enforcing fleet-level caps on CO$_2$ and energy use, and respecting per-vehicle operating limits. The optimization is solved using the HiGHS linear programming solver via SciPy’s linprog. Results show that shifting utilization away from high-intensity vehicles and toward more efficient units yields substantial system-level improvements with no reduction in total operating hours: the optimized schedule achieves an approximately 31\% reduction in total CO$_2$eq and a simultaneous 9.4\% reduction in total energy consumption.


\end{abstract}

\section{Introduction}


Maritime transport remains the backbone of international trade—moving over 80\% of goods traded worldwide by volume—which makes ports critical system nodes whose operational efficiency has direct implications for energy use and environmental performance \cite{UNCTAD}.  Alongside the broader push for sustainable industry—reflected in initiatives such as the European Green Deal and the Sustainable Development Goals (SDGs)—there is growing pressure to reduce the carbon footprint of maritime transport and port‑related operations without compromising throughput.  Recent evidence shows that emissions from global shipping remain substantial and have continued to rise in the post‑pandemic period: the OECD’s AIS‑based estimates indicate that global maritime transport CO$_2$ emissions increased from 889 Mt (2019) to 973 Mt (2024) (+9.4\%), with a marked acceleration in 2024 \cite{de2025analysing}.  Independent inventory work similarly reports large and persistent shipping emissions through 2016–2023, reinforcing the urgency of operational measures that can cut emissions in the near term \cite{mao2025greenhouse}.  Because ports are repeatedly identified as strategic intervention points for decarbonisation—through operational efficiency, electrification, and smarter energy use—optimization of day‑to‑day logistics decisions (e.g., fleet scheduling and utilization) becomes a practical lever for measurable emissions reduction within existing infrastructure constraints.

To act on these decarbonisation pressures in a measurable and operationally feasible way, ports increasingly rely on Internet of Things (IoT)--enabled digitalization to transform physical processes into data streams that can be monitored, analysed, and optimized in near real time \cite{yang2018internet, bourgioukou2023investigating,hussain2026digital}. Smart-port research emphasizes that IoT infrastructures provide the visibility layer needed to coordinate complex, multi-actor terminal operations (e.g., equipment status, asset condition, and logistics flows), thereby enabling data-driven decisions that improve efficiency and sustainability outcomes \cite{yau2020towards}. Recent literature reviews further highlight digitalization as a defining characteristic of smart ports, supporting automation, energy management systems (EMS), and performance improvement by integrating sensing, connectivity, and data platforms into operational workflows \cite{sadiq2021future,  yang2025assessing, heilig2017information}. From an energy-and-emissions perspective, seaport decarbonisation studies similarly identify digital technologies as key enablers because they strengthen situational awareness and make operational optimisation actions—such as scheduling, resource allocation, and energy-aware planning—implementable at scale \cite{alzahrani2021decarbonisation}.

In the above context, the digital visibility enabled by smart-port IoT infrastructures, the present work leverages operational data a real-world case study at the Port of HaminaKotka (Mussalo terminal) to demonstrate a scheduling approach that targets CO$_2$eq reduction in landside fleet operations. Landside vehicle fleets represent a controllable component of port operations, and prior studies have highlighted fleet scheduling and resource allocation as promising levers for improving efficiency and environmental performance in logistics and industrial systems \cite{chhikara2020efficient, kim2022study, zhong2023energy, zhen2018cranes}. 
Using a two-year dataset from 77 port-operating vehicles, we formulate the scheduling task as a linear programming–based workload allocation problem that reallocates operating hours to minimize total CO$_2$eq emissions while maintaining fixed operational demand and respecting feasible operating limits. Linear optimization has been widely adopted for large-scale scheduling and energy-aware resource allocation due to its transparency, computational efficiency, and suitability for decision support in industrial environments \cite{bertsimas1997introduction}. By grounding the formulation in operational port data, the proposed approach demonstrates how IoT-enabled digital processes can be coupled with optimization to deliver practical, near-term decarbonization of port logistics without requiring changes to infrastructure or service levels. The key contributions of the work are summarized below.
\begin{itemize}
    \item The paper provides an overview of  IoT-enabled port digitalization with optimization by showing how operational data generated through digitalized port processes can be transformed into actionable scheduling decisions for emissions reduction in landside logistics.
\item A linear programming–based workload allocation model is developed and applied to a real-world case study at the Port of HaminaKotka (Mussalo terminal) using a two-year dataset from 77 vehicles, minimizing total CO$_2$eq emissions while preserving total workload and respecting fleet-level and per-vehicle constraints.

\item The proposed scheduling approach achieves an approximately 31\% reduction in total CO$_2$eq emissions and a 9.4\% reduction in energy consumption solely through schedule reallocation, demonstrating a practical, near-term decarbonization lever that requires no infrastructure changes.
    
\end{itemize}
The rest of the paper is organized as follows: Section II provides key enabling technologies for an IoT-enabled framework, Section III formulates the fleet scheduling problem, Section IV explains the results and discussion, and finally, Section V provides the conclusion of the paper.

\section{Key Enabling Technologies for an IoT‑Enabled Data‑to‑Decision Framework for Port Decarbonization}

\subsection{IoT‑Based Operational Data Acquisition and Integration}

IoT devices assemble and integrate spatial and temporal data
such as humidity, temperature, and the physical location of the objects \cite{yau2020towards}. This data is sourced from various heterogeneous sensing, monitoring, and detecting devices,
including cameras, sensors, radio frequency identification (RFID) readers and tags, face recognition systems, and bar codes. These devices can be equipped at various port areas, for instance, in qubay, terminal, yard, warehouses– and infrastructural locations (i.e., quay cranes, yard storage, and container terminals) to keep track of environmental conditions and cargo schedules and share the information with the data centers \cite{heilig2017information}. The integration of IoT‑based data acquisition improves visibility of equipment usage, fleet activity, and operational states, forming the foundation for data‑driven optimization and automation.  In the HaminaKotka context, IoT-enabled sensing and monitoring technologies, such as GISGRO are part of comprehensive smart port initiatives \cite{Digi, NUS, GISGRO_1}. GISGRO platform consolidates inspection and structural data, encompassing both underwater and above-ground measurements, thereby facilitating maintenance planning across the port’s infrastructure. It effectively operates as a structural health information system for the port. The Finnish smart ports have demonstrated the use of mobile networks, including trials with narrowband IoT within 5G testbeds, to facilitate IoT sensor connectivity in logistics, situational awareness, and automation projects \cite{Xamk}. This suggests that similar low-power wide-area IoT technologies, such as NB-IoT and LTE-M, could be effectively
utilized for analogous sensing applications within the HaminaKotka ecosystem.

\begin{figure}
    \centering
    \includegraphics[width=0.5\textwidth]{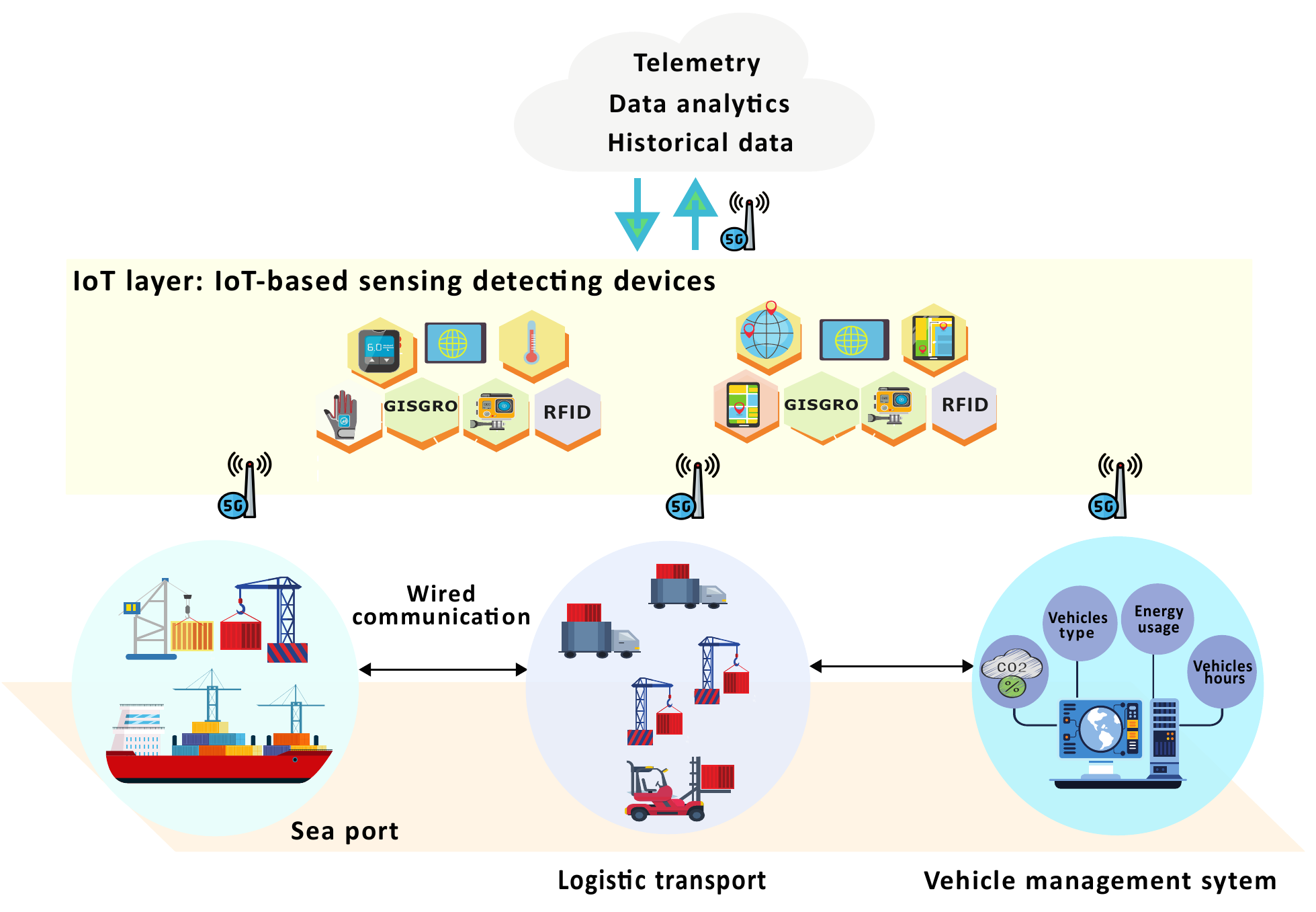}
    \caption{System-level overview of the designed framework.}
    \label{T}
    \vspace{-0.2cm}
\end{figure}

\subsection{Advanced Communication and Edge–Cloud Computing Infrastructure}
Smart‑port digitalization relies on communication and computing infrastructures that can handle large volumes of operational data with sufficient reliability and scalability. Edge–cloud architectures are widely discussed in the literature as a practical approach for industrial IoT systems, balancing local responsiveness with centralized data management \cite{pahl2015containers, belcastro2026navigating}.
In the Port of HaminaKotka, and specifically at the Mussalo terminal, advanced communication infrastructure has been introduced to support digitalized container and logistics operations through private cellular networking \cite{NUS, Digi}. Port has private LTE networks to connect cranes, vehicles, cameras, and operational systems, enabling continuous data exchange across the terminal and supporting real‑time operational visibility and control. More recently, the introduction of private 5G connectivity at Mussalo further strengthens this communication layer by providing higher capacity and lower latency for data‑intensive applications such as video streams, telemetry, and production‑system communication \cite{Digi_1}.  Computing support is limited to centralized data handling for analysis and decision support, while operational control remains at the field level  \cite{Digi}. This setup is sufficient to enable data availability and data-driven applications, which results the basis for emissions‑aware fleet scheduling and optimization in this study.  

\section{System description}
Landside vehicle fleets at the Port of HaminaKotka (Mussalo terminal) are essential to container‑handling operations and contribute significantly to operational energy use and CO$_2$eq emissions.  As depicted in Fig.  \ref{T}, within the context of a seaport environment, vehicles are tasked with the transportation of cargo from maritime vessels to container terminal (CT) and from CT to trailers for outbound logistics. In this operational framework, the vehicle management system is integral to the efficient scheduling of these activities, thereby contributing to the reduction of CO$_2$ equivalent emissions. We consider a fleet of $N$ port-operating vehicles indexed by $i \in \{1, 2, \ldots, N\}$. Each vehicle activity  is digitally recorded through IoT-based operational data acquisition and integration using terminal operating system (TOS), vehicle logs, automatic
vehicle identification systems (e.g., RFID and gate-based OCR/ANPR), and metered records of operating hours and energy consumption.  
Each vehicle performs logistics tasks such as container handling and material transport during the planning horizon. The dataset consists of aggregated operational records from 77 container-handling vehicles operating at the Port of HaminaKotka over a two-year observation period. All energy consumption and emission values reported in this study represent cumulative fleet-level quantities obtained directly from the port database. Consequently, the magnitudes reported in the results section correspond to aggregated fleet operations over the complete study period rather than individual vehicle or annual values
Based on historical operational data records, each vehicle is characterized by: (i) an average CO$_2$eq emission rate $e_i$ (kg CO$_2$eq per hour), and  (ii) maximum feasible number of operating hours $H_i^{\text{max}}$, derived from observed availability and utilization patterns. These data are transmitted via advanced communication infrastructure to centralized port information systems for aggregation and decision support, while operational control remains at the field level. Although total workload is fixed by port demand, the distribution of work across heterogeneous vehicles is flexible, creating an opportunity to reduce fleet‑level CO$_2$eq emissions through improved scheduling. This study therefore addresses how operating hours should be allocated across the existing fleet to minimize total CO$_2$eq emissions while preserving throughput and respecting practical operating limits using digitally available operational and energy data.

\begin{figure*}[!t]
\centering
\setlength{\abovecaptionskip}{2pt}
\setlength{\belowcaptionskip}{0pt}

\begin{minipage}[b]{0.45\textwidth}
    \centering
    \includegraphics[width=\textwidth, height=0.25\textheight, keepaspectratio]
    {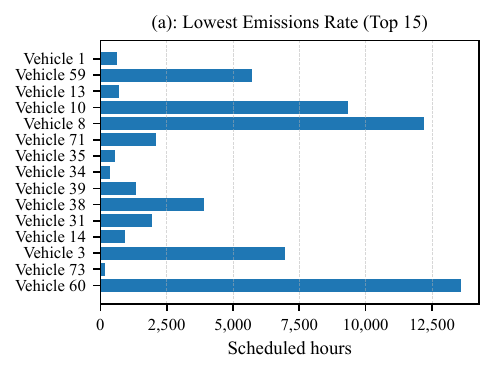}
\end{minipage}
\hspace{0.02\textwidth}
\begin{minipage}[b]{0.45\textwidth}
    \centering
    \includegraphics[width=\textwidth, height=0.25\textheight, keepaspectratio]
    {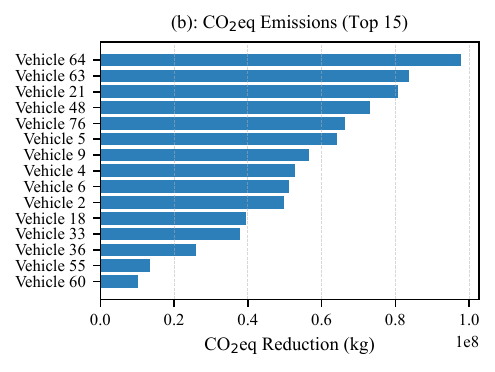}
\end{minipage}
\caption{(a) Top 15 vehicles with lowest CO$_2$eq emission rates over the two-year study period and (b) Top 15 vehicles with highest cumulative CO$_2$eq reductions over the two-year study period.}
\label{fig:emission_combined}

\vspace{-0.4cm}
\end{figure*}

\section{Problem formulation}
\vspace{-0.2cm}
This section formulates the emissions-aware fleet scheduling problem for landside operations at the Port of HaminaKotka (Mussalo terminal). The objective is to optimally allocate operating hours across a heterogeneous fleet of port vehicles in order to minimize total CO$_2$eq emissions, while maintaining the same overall operational workload and respecting practical operating limits. The total required workload over the planning horizon is assumed to be fixed and equal to the historically observed total operating hours. Based on this the objective function is formulated to to minimize the total fleet‑level CO$_2$eq emissions resulting from vehicle operation:
{\setlength{\abovedisplayskip}{2pt}
\setlength{\belowdisplayskip}{0pt}
\begin{align}
\min \quad & Z = \sum_{i=1}^{n} e_i h_i
\label{Obe}
\end{align}}
\eqref{Obe} prioritizes the allocation of work toward lower‑emission vehicles while discouraging the use of higher‑intensity units. To ensure operational feasibility, the optimization is subject to the following constraints.
{\setlength{\abovedisplayskip}{2pt}
\setlength{\belowdisplayskip}{2pt}
\begin{align}
\text{s.t.} \quad
& \sum_{i=1}^{n} h_i = H_{\text{total}} \label{eq1} \\
& 0 \le h_i \le H_i^{\text{max}} \label{eq2}\\
& \sum_{i=1}^{n} e_i h_i \le C_{\text{max}} \label{eq3}\\
& \sum_{i=1}^{n} k_i h_i \le E_{\text{max}} \label{eq4}
\end{align}

Where, \eqref{eq1} is workload preservation constraint and  $H_{\text{total}}$ denotes the total required operating hours, computed from historical data, \eqref{eq2} is vehicle operating limits and constraints \eqref{eq3} and \eqref{eq4} are fleet‑level emissions and energy caps and $ C_{\text{max}}$ $E_{\text{max}}$ represent allowable upper bounds on total  CO$_2$eq emissions and total energy use, respectively, and $ k_i$ denotes the energy intensity of vehicle $i$. The resulting optimization problem is a linear program (LP) and is solved using the HiGHS solver via SciPy’s linprog interface. HiGHS is a state‑of‑the‑art open‑source LP solver designed for large‑scale sparse problems and provides reliable performance and numerical robustness for industrial scheduling applications. The LP structure allows the solution to be obtained efficiently, enabling rapid evaluation of alternative workload allocations without changes to model formulation. In this study, the solver computes the emissions‑minimizing allocation of operating hours subject to workload preservation, fleet‑level emissions and energy caps, and per‑vehicle operating limits.

\section{Results and dicussion}
In this section, we present the findings and discussion regarding the solution methodologies employed to address the vehicle missions-aware fleet scheduling problem pertinent to landside operations at the Port of HaminaKotka, specifically focusing on the Mussalo terminal. This analysis is grounded in a real-world case study utilizing a comprehensive two-year dataset encompassing 77 vehicles operated. To ensure data confidentiality, each vehicle is assigned a fixed anonymous label (Vehicle 1, Vehicle 2, $\dots$), consistently used across all results and figures. The proposed scheduling framework incorporates vehicle-specific intensity metrics, specifically CO$_2$eq per operating hour, alongside energy consumption considerations.

Emissions are recorded directly as cumulative CO CO$_2$eq emissions (kg  CO$_2$eqeq) for each vehicle. Energy consumption values originate from the port operational database and are stored separately from the derived energyTJ field. All results reported in this paper correspond to fleet-level cumulative values aggregated over the complete study period rather than annual or per-vehicle values

\begin{figure*}[!t]
\centering
\setlength{\abovecaptionskip}{2pt}
\setlength{\belowcaptionskip}{0pt}

\begin{minipage}[b]{0.45\textwidth}
    \centering
    \includegraphics[width=\textwidth, height=0.25\textheight, keepaspectratio]
    {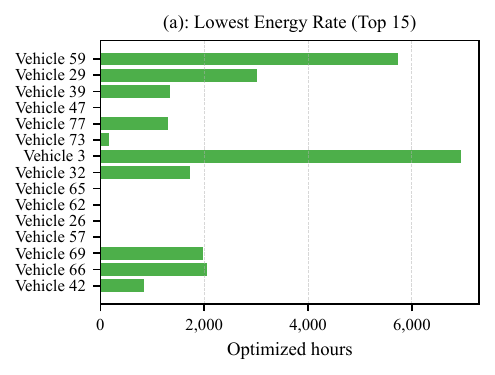}
\end{minipage}
\hspace{0.02\textwidth}
\begin{minipage}[b]{0.45\textwidth}
    \centering
    \includegraphics[width=\textwidth, height=0.25\textheight, keepaspectratio]
    {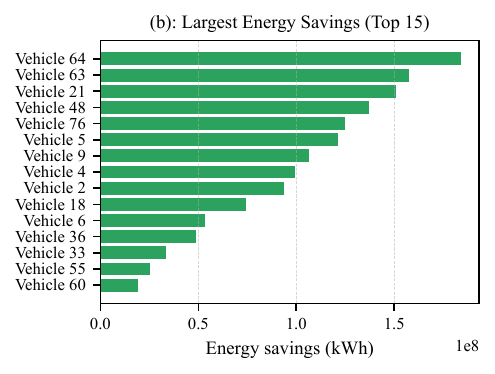}
\end{minipage}
\caption{(a)  Top 15 vehicles with lowest energy usage over the two-year study period and (b) with the highest energy reductions over the two-year study period.}
\label{fig:energy_combined}

\vspace{-0.4cm}
\end{figure*}

Fig. ~\ref{fig:emission_combined}(a) presents the optimized scheduled hours for the 15 vehicles that exhibit the lowest emissions (kg CO$2$eq/h), where efficiency is assumed as the reduction of emissions per unit of time.  The model operates under the constraint of a fixed total workload, denoted as in equation \eqref{eq2} and simultaneously enforcing an emissions cap by equation \eqref{eq4}. Consequently, operational hours are preferentially allocated to these low-emission vehicles, thereby enabling the completion of the predefined workload with diminished emissions. The differences in allocated hours can be attributed to capacity constraints, specifically the condition that the hours allocated to each vehicle must not exceed 1.5 times their respective baseline hours. Vehicle 1 is assigned fewer hours than Vehicle 60, a situation that arises from its lower baseline, which limits the potential for an increase in allocated hours. In contrast, Vehicle 60 is able to accommodate a greater number of hours owing to its higher baseline. Meanwhile, Vehicle 73 receives the least allocation, due to its restricted baseline utilization.

Fig. ~\ref{fig:emission_combined}(b) shows the vehicles with the highest absolute reductions in total CO$_2$eq after optimization. The largest savings come from high‑emission, high‑utilization units whose hours were cut sharply or even reduced to zero. These vehicles contribute disproportionally to the baseline total, so cutting their hours yields large reductions in kg CO$_2$eq. In practice, this group is dominated by the highest‑rate heavy units (e.g., Vehicle fleet  64, 63, 21, 55, and 60), which explains why their bars are the largest in this savings plot.

Importantly, a vehicle can be efficient per hour yet still show small savings here if it receives more hours (because its low rate makes it attractive for reallocation). Therefore, Fig.  ~\ref{fig:emission_combined}(b) reflects where the system can remove the most emissions, not necessarily which vehicles are the most efficient per hour. Together, Fig. ~\ref{fig:emission_combined}(a) and ~\ref{fig:emission_combined}(b) show the same mechanism from two sides: hours shift toward low‑rate vehicles and away from high‑rate vehicles, producing the largest CO$_2$eq reductions among the latter.

Fig. ~\ref{fig:energy_combined}(a) demonstrates the emissions with energy intensity (kWh/h). Here, best performing means lowest kWh per hour, so the optimizer shifts hours toward those vehicles to keep total work fixed while staying under the energy cap ($\sum_i e_i h_i \le E_{\max}$), where $E_{\max}$ is the current total energy baseline. In low‑rate group, bar heights: vehicles with zero optimized hours (e.g.,  Vehicle 57, 26, 62, 65, and 47) can still be energy‑efficient, but their baseline hours are very small or the optimizer can meet total demand using other low‑rate vehicles within constraints, so their optimized hours stay at zero. Vehicles with very high optimized hours e.g., Vehicle 59, 29, 3, 69, and 66  started with large baseline utilization and therefore, have much higher allowable upper bounds, letting them absorb more reallocated workload while keeping energy use low per hour. Vehicles 39, 77, 73, 32, and 42 are still energy efficient, but they are only partially used because their capacity is smaller and the total hours target is already met by other low‑rate vehicles. Fig. ~\ref{fig:energy_combined}(b) depicts vehicles that deliver reductions in total energy (kWh) after optimization. The highest savings come from energy-intensive, heavily used vehicles whose hours were cut substantially. Efficient vehicles can show smaller savings or even negative savings if they absorb more hours, because the system reallocates work to minimize total energy while keeping the same total hours. From Fig. ~\ref{fig:energy_combined}(a) and ~\ref{fig:energy_combined}(b), it can be noticed that the optimization shifts hours away from high‑energy‑intensity units and toward low‑energy‑intensity ones.

\begin{figure}
    \centering
    \includegraphics
    [width=\columnwidth, trim=5 5 5 5, clip]{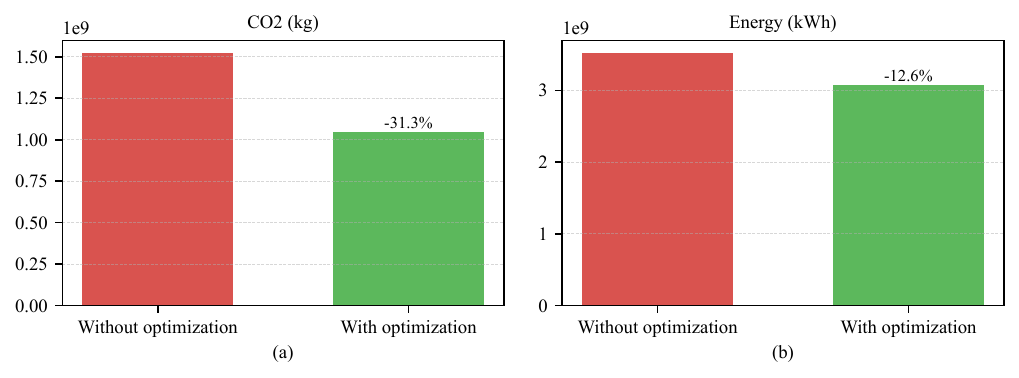}
    \caption{Total (a) CO$_2$eq and (b) energy with and without optimization}
    \label{Total_Re}
    \vspace{-0.2cm}
\end{figure}
Note that the reported CO$_2$eq and energy values correspond to cumulative fleet-level totals aggregated across 77 vehicles over the complete two-year study period.
Fig. ~\ref{Total_Re} reflects the total CO$_2$eq and energy before vs. after optimization for CO$_2$eq and energy (kWh). The optimized scheduling framework maintains the total work hours while redistributing these hours among various vehicles to achieve a reduction in system emissions, all while adhering to established energy and capacity constraints. Consequently, the total CO$_2$eq emissions and energy consumption are reduced to 31.26\% and 12.64\%, respectively. Note that, the optimal allocation of operational hours from vehicles with high emissions and elevated energy consumption to those with lower emissions and energy usage rates, all while remaining within the specified upper limits for each vehicle. This approach effectively redistributes the workload to cleaner and more energy-efficient vehicles without compromising the overall demand for services. The detials of the Optimization summary is presented in Table \ref{T1}.

\begin{table}[!t]
\caption{Optimization summary}
\label{T1}
\centering

\begin{threeparttable}

\begin{tabular}{lr}
\toprule
\textbf{Parameters} & \textbf{Value} \\
\midrule
Vehicles & 77 \\
Total work hours & 238,288 \\
Current CO$_2$eq (kg) & 1,521,963,883 \\
Optimized CO$_2$eq (kg) & 1,046,218,185 \\
Total CO$_2$eq reduction (kg) & 475,745,698 \\
Total CO$_2$eq reduction (\%) & 31.3\% \\
Current energy (kWh) & 3,687,122,482 \\
Optimized energy (kWh) & 3,342,061,774 \\
Total energy reduction (kWh) & 345,060,708 \\
Total energy reduction (\%) & 9.4\% \\
\bottomrule
\end{tabular}

\begin{tablenotes}
\footnotesize
\item Note: Results represent cumulative totals aggregated across all 77 vehicles during the two-year study period.
\end{tablenotes}

\end{threeparttable}

\end{table}

\section{Conclusion}

A linear programming framework was developed and applied to allocate vehicle work hours in a manner that minimizes CO$_2$eq emissions while honoring fixed total demand and energy limits. By solving a compact LP with the HiGHS solver, the method efficiently reassigns workload toward the most efficient vehicles and away from high‑emission units. The case results demonstrate that significant environmental benefits can be achieved without curtailing operations; in this study, total CO$_2$eq dropped by 31.26\% , accompanied by 12.64\% reduction in total energy use. The vehicle‑level shift analysis further clarifies the mechanism of improvement: high‑intensity vehicles experience sharp reductions in hours, whereas low‑intensity vehicles take on additional utilization to satisfy demand. These findings indicate that optimization‑based scheduling is an effective, explainable tool for fleet decarbonization. Future work should integrate operational realism through minimum/maximum utilization constraints, vehicle availability and maintenance windows, and multi‑objective formulations that jointly optimize emissions, energy, and cost.

\section*{Acknowledgment}
The studies were made in the project Smart grid and energy community pilot in Mussalo, Kotka funded by the EU and managed by Regional Council of Kymenlaakso Finland, funding decision EURA 2021/902692/09.

\bibliographystyle{ieeetr}
\bibliography{Ref}
\end{document}